\documentclass[fleqn,10pt]{wlscirep}
\usepackage{braket}
\usepackage{graphicx}
\usepackage{siunitx}
\usepackage{amsmath}

\begin{document}

\title{A Study on Fast Gates for Large-Scale Quantum Simulation with Trapped Ions}

\author[1,*]{Richard L. Taylor}
\author[1,$\dagger$]{Christopher D. B. Bentley}
\author[2]{Julen S. Pedernales}
\author[2]{Lucas~Lamata}
\author[2,3]{Enrique Solano}
\author[1,4,$\ddagger$]{Andr\'e R. R. Carvalho}
\author[1]{Joseph J. Hope}
\affil[1]{Department of Quantum Science, Research School of Physics and Engineering, Australian~National~University, Canberra ACT 0200, Australia}
\affil[2]{Department of Physical Chemistry, University of the Basque Country UPV/EHU, Apartado~644, Bilbao 48080,
Spain}
\affil[3]{IKERBASQUE, Basque Foundation for Science, Maria Diaz de Haro 3, 48013 Bilbao, Spain}
\affil[4]{ARC Centre for Quantum Computation and Communication Technology, Australian~National~University, Canberra ACT 0200, Australia}
\affil[*]{email: u5471774@anu.edu.au}
\affil[$\dagger$]{Current address: Max-Planck-Institut f\"ur Physik komplexer Systeme, N\"othnitzer Str. 38, D-01187 Dresden, Germany}
\affil[$\ddagger$]{Current address: Centre for Quantum Dynamics, Griffith University, Nathan QLD 4111, Australia}

\begin{abstract}
Large-scale digital quantum simulations require thousands of fundamental entangling gates to construct the simulated dynamics.
Despite success in a variety of small-scale simulations, quantum information processing platforms have hitherto failed to demonstrate the combination of precise control and scalability required to systematically outmatch classical simulators.
We analyse how fast gates could enable trapped-ion quantum processors to achieve the requisite scalability to outperform classical computers without error correction.
We analyze the performance of a large-scale digital simulator, and find that fidelity of around 70\% is realizable for $\pi$-pulse infidelities below $10^{-5}$ in traps subject to realistic rates of heating and dephasing.
This scalability relies on fast gates: entangling gates faster than the trap period.
\end{abstract}

\flushbottom
\maketitle

\thispagestyle{empty}

\noindent 

\section*{Introduction}

Quantum simulation promises the ability to study the dynamics of highly complex quantum systems using more easily accessible and controllable systems~\cite{Feynman1982,Buluta2009,Georgescu2014}. Digital quantum simulation schemes, using quantum computing resources, enable versatile simulations of a collection of systems - indeed, a universal set of quantum gates is sufficient to compose any desired unitary arising from a local Hamiltonian~\cite{Lloyd1996}. In contrast, analogue quantum simulation schemes require precise engineering of a Hamiltonian to reproduce the desired dynamics of the simulated system~\cite{Georgescu2014}, such that they are necessarily less versatile.  Moreover, digital simulators may allow for error correction due to their similarities with gate-based quantum computing schemes.

Many systems are considered as potential quantum simulators: cold atoms in optical lattices~\cite{Bloch2012}, superconducting circuits~\cite{You2011}, and nuclear spin systems~\cite{Zhang2012,Peng2009}, in addition to trapped ions \cite{Blatt2008,Wineland1998}. 
Each simulator platform has its own advantages and challenges - cold atoms scale well but are difficult to control individually, whereas trapped ions and superconducting circuits face scaling difficulties but have experimentally-demonstrated individual control and readout techniques~\cite{Haffner2008}. 
Both trapped ions~\cite{Lanyon2011,Martinez2016,Barreiro2011,Casanova2012,Mezzacapo2012,Lamata2013} and superconducting circuits~\cite{Salathe2015,Barends2015a,Barends2015b} have demonstrated great potential for implementing digital quantum simulations, using to date up to nine qubits.
Here we investigate the feasibility of a digital simulation scheme with trapped ions. This scheme\cite{Lamata2013} proposes a digital simulation of fermionic Hamiltonians using M\o lmer-S\o rensen gates in an ion trap.

In this work, we show that a simpler gate, equally capable of efficiently implementing the fermionic simulation, may be constructed using faster resonant two-qubit gates \cite{Garcia-Ripoll2003}. We also show that M\o lmer-S\o rensen gates are incapable of implementing the simulation at a sufficient scale to outperform classical computers within the coherence time of current ion traps, and that fast gates could enable this level of performance using near-future laser technology. Our results motivate the pursuit of fast gates as a critical scaling tool for trapped-ion computing architectures.

\subsection*{Simulation scheme}
We investigate a digital quantum simulation of a fermionic lattice system using an ion trap \cite{Lamata2013}. A Jordan-Wigner transformation maps a lattice fermionic Hamiltonian to a qubit Hamiltonian, and then a Trotter-Suzuki expansion can be used to decompose the resulting evolution into discrete steps. With an ion trap, a digital simulation of this system can be implemented efficiently using steps of the form
\begin{equation}
\label{Unitarydecomposition}
U_\textrm{MS}^\dagger\left(\frac{\pi}{2}\right)U_{\sigma_z^m}U_\textrm{MS}\left(\frac{\pi}{2}\right) .
\end{equation}



Here, $U_\textrm{MS}(\frac{\pi}{2})$ is an MS gate and $U_{\sigma_z^m}$ is a local rotation on the $m$th ion in the trap. The MS gate couples every pair of ions in the trap evenly,
\begin{equation}
\label{MSgate}
U_\textrm{MS}(\theta)=e^{-i\frac{\theta}{4}\sum_{k,l=1}^L\sigma_x^k\sigma_x^l},
\end{equation}

where $\sigma_x^k$ is the Pauli $x$ operator on ion $k$, and $L$ the total number of ions.
Although MS gates can be implemented with high fidelity in ion traps~\cite{Benhelm2008, Monz2011}, they are much slower than the trap period due to their reliance on the Lamb-Dicke regime and the vibrational rotating-wave approximation.
Furthermore, as the number of addressed ions increases to scale up the simulation, the motional sideband spectral density increases and the gate time increases accordingly. 
It would thus be difficult to use MS gates, or alternative sideband-resolving gates, for a digital quantum simulation of a sufficient size to outperform current classical computers, with dozens of interacting spins or fermions.

It was also shown in the simulation proposal\cite{Lamata2013} that each MS gate can be replaced by an `ultrafast multiqubit' (UMQ) gate in the simulation step,
\begin{equation}
U_\textrm{UMQ}=e^{-i\frac{\pi}{4}\sigma_x^m\sum_{k=1,k\neq m}^L\sigma_x^k} ,\label{eq3}
\end{equation}
coupling only pairs involving the locally-rotated ion.
It was suggested that these UMQ gates could be constructed using fast two-qubit phase gates~\cite{Garcia-Ripoll2003}, which can be performed much faster than the trap period on ion crystals of arbitrary length, with high fidelity~\cite{Bentley2015,Duan04PRL,Zhu06EL,GZC05PRA}. 
In light of recent progress towards the implementation of the first fast entangling gate using ultrafast laser pulses~\cite{Camp10PRL,Bentley2013,Mizr13PRL,Mizr14APB,Huss16pc}, we explore this approach to validate the pursuit of fast gates as a scaling tool as well as to benchmark the required laser control and environmental coupling rates.

In this paper, we perform numerical simulations of fast gates in a full 40-qubit crystal to calculate the fidelity of a digital quantum simulation scheme under typical noise parameters.
We benchmark the laser repetition rates which will allow us to overcome the time barrier imposed by the MS gate and significantly reduce the operation time of digital quantum simulations, as well as greatly increasing the final fidelity. 
The latter implies that large-scale quantum simulations could be performed without error correction, bringing nearer the solutions for problems in condensed matter, quantum chemistry, and high-energy physics, which are infeasible to classical computers. 

\section*{Results}

\subsection*{Fast gate model}

Fast two-qubit gates can be performed beyond the trap-period limit and implement the ideal unitary for a geometric phase gate,
\begin{equation}\label{ideal}
U_\textrm{I} = e^{i\frac{\pi}{4}\sigma_z^1\sigma_z^2},
\end{equation}
where $\sigma_z^k$ is the Pauli $z$ operator on the $k$th ion. These gates are implemented using counter-propagating laser $\pi$-pulse pairs, which act as state-dependent momentum kicks on the motional modes while preserving the internal states of the ions. This behaviour is described in the optical rotating-wave approximation by the kick unitary
\begin{equation}
U_\textrm{kick}=e^{-2i\zeta k(x_1\sigma_z^1+x_2\sigma_z^2)}.
\end{equation}
Here, $x_c$ is the position operator on ion $c$, $\zeta$ is the number of $\pi$-pulse pairs used for the momentum kick and $k$ is the laser wavenumber. These kicks are nearly instantaneous compared to the trap period, so the evolution of a real gate operation can be expressed as the product of the kick unitaries interspersed with free motional evolution,
\begin{equation}
U_\textrm{re}=\prod_{c=1}^N\left(\prod_{p=1}^Le^{-i\nu_pa^\dagger_pa_p\delta t_c}\right)U^c_\textrm{kick},
\end{equation}
where $U^c_\textrm{kick}$ is the $c$th kick unitary, $\delta t_c$ is the time between kicks $c$ and $c+1$ (0 if $c=N$) and $N$ is the total number of kicks. An analytic solution for this evolution to implement the ideal unitary in Eq.~(\ref{ideal}) cannot generally be found, so numerical methods are used to optimise the fidelity of the resulting gate. The high-fidelity solutions found by this optimisation approach will restore the initial motional state of the trap, while the state-dependent phase-space trajectories for the motional modes will enclose areas corresponding to the $\frac{\pi}{4}$ phase required by the ideal unitary. This results in a geometric phase gate that produces minimal trap heating.

The structure of the kicks is chosen to be antisymmetric, which approximately restores the motional state - the dominant motional error term is cubic with respect to gate time \cite{Bentley2015}. This reduces the difficulty of the numerical optimisation and produces generally higher-fidelity solutions. In particular, the Fast Robust Antisymmetric Gate (FRAG) scheme~\cite{Bentley2013} was found to give the shortest gate times for a target fidelity using realistic experimental parameters. This scheme uses the kick sequence
\begin{align}
\zeta &= (-n,2n,-2n,2n,-2n,n), \nonumber \\
t &= (-\tau_1,-\tau_2,-\tau_3,\tau_3,\tau_2,\tau_1),
\end{align}
where $z$ is the set of kick magnitudes associated with the times $t$ and $n$ is an integer. Negative and positive kick amplitudes represent kicks in opposite directions. The originally proposed FRAG scheme restricts the order of the kicks, so $\tau_1>\tau_2>\tau_3>0$. Removing this restriction preserves the antisymmetric structure of the sequence whilst allowing numerical optimisations to search a larger parameter space. This often results in higher-fidelity gate solutions, so we ignore the kick ordering restriction in this work.

We address the axial phonon modes of the trap - it is possible to perform quantum gates using transverse phonon modes \cite{Zhu2006}, but for a typical linear trap the electrode geometry will limit their speed due to micromotion. Addressing the axial modes requires that ground-state population in the ions not addressed by each fast gate be `shelved' \cite{Leib03RMP} into a third state such that they are not affected by the fast gate pulses. In the following analysis we neglect the infidelity of the shelving technique, as each ion can be shelved with a single optical $\pi$ pulse (albeit at a different wavelength) and thus errors will contribute negligibly compared with the infidelity of the fast gates, which use hundreds of $\pi$ pulses.

We can use fast geometric phase gates to implement the UMQ gates required in digital quantum simulations. If we label the ions in a Paul trap from 1 to $L$ starting at one end, this is done by performing a fast gate on ions 1 and 2, a SWAP gate on these ions, then a fast gate on ions 2 and 3 and so on until we reach ions $L-1$ and $L$. The SWAP gate is given in the computational basis by
\begin{equation}
U_\textrm{SWAP}=\begin{pmatrix}1&0&0&0\\0&0&1&0\\0&1&0&0\\0&0&0&1 \end{pmatrix},
\end{equation}
and is equivalent to `swapping' the internal states of the two ions.

This process gives the UMQ unitary required up to local rotations, not preserving the position of the internal states in the trap. The reversed UMQ gate in the simulation step restores the position of the internal states. It should be noted that this method of constructing a UMQ gate requires fast SWAP gates, which can be implemented with three CNOT gates \cite{Nielsen2010}. Each CNOT requires one fast geometric phase gate and four local rotations.

This allows us to implement a non-position-preserving UMQ gate using $L-2$ SWAP gates and an additional $L-1$ fast gates, for a total of $4L-7$ fast gates. Due to the relative timescales and fidelities of local rotations, which require only single laser pulses that take negligible time and can be performed with very high fidelity~\cite{Harty2014}, we consider only infidelity and time cost of the fast two-qubit gates in our results. This approximation is valid because the fast two-qubit gates require hundreds of laser pulses each, so will contribute far more error and take far more time than the few single-pulse local rotations required in the algorithm.

\subsection*{Simulation fidelity calculations}

We estimated fidelity $F$ for a UMQ gate by first calculating the fidelities of the two-qubit fast gates used to construct the UMQ operation. This calculation was performed with a state-averaged fidelity measure, assuming an initial thermal product state for the motional modes of the trap. This fidelity measure is given by
\begin{eqnarray}
F \! = \! \frac{ \int_{\psi_0} \!\! \textrm{Tr}_m{[}\left<\psi_0\right|U_\textrm{id}^\dagger U_\textrm{re}\left|\psi_0\right>\left<\psi_0\right|\otimes \rho_m U_\textrm{re}^\dagger U_\textrm{id}\left|\psi_0\right>{]}\ \!\! \textrm{d}\ket{\psi_0} }{\int_{\psi_0}\textrm{d}\ket{\psi_0} },
\end{eqnarray}
where $U_{\rm re}$ and $U_{\rm id}$ are the real and ideal gate operations, respectively. The assumption of a thermal product state simplifies expectation values of motional displacement operators that appear upon expanding this expression, allowing the calculation of an analytic expression for the state-averaged fidelity given a set of kick times. This expression simplifies numerical optimisation by removing the need to calculate the full unitary $U_\textrm{re}$. We optimised the component gate fidelities for every adjacent pair of ions in Paul traps of various sizes. The longitudinal trap frequency was chosen as $\nu=2.1856 \textrm{ MHz} \times L^{-0.865}$, where the power of $L$ represents a no-buckling limit for the trap derived from molecular dynamics simulations \cite{Wineland1998a,Schiffer1993} and the constant coefficient is chosen to give a typical trapping frequency of 1.2 MHz for a two-ion trap. The transverse trap frequency was constant at $\nu_x=5$ MHz. The Lamb-Dicke parameter was $\eta=0.16$ for the two-ion trap, and scaled appropriately with trapping frequency for larger traps~\cite{Benhelm2008}.
\begin{figure}[h]
\includegraphics[width=1\textwidth]{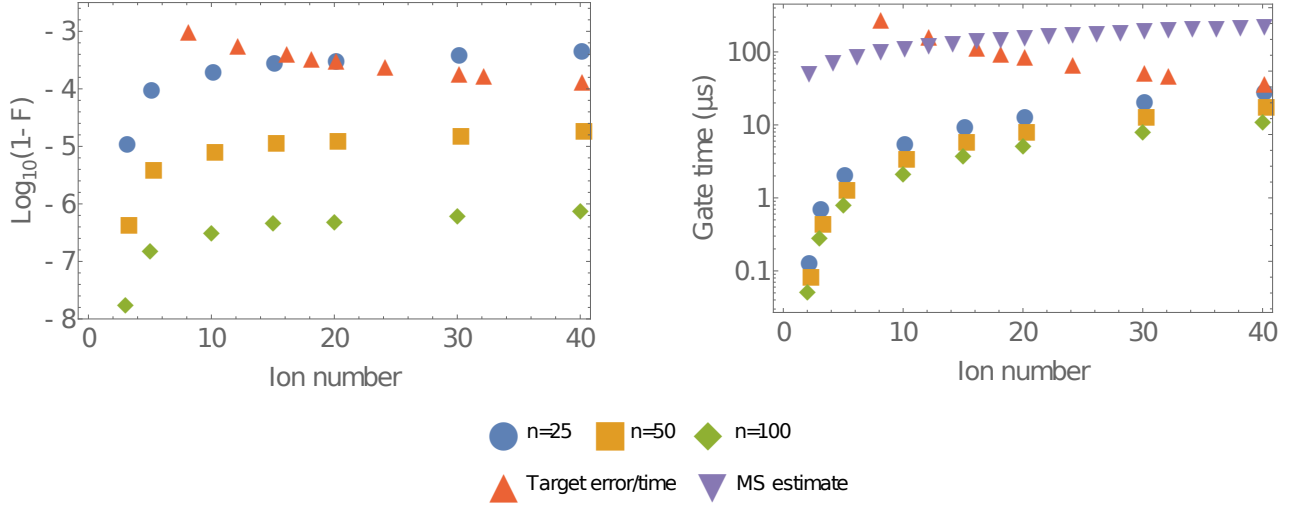}
\caption{Plots of (a) error and (b) total time taken for fast-gate-based algorithms implementing the UMQ unitary (Equation \ref{eq3}). $n$ represents the number of $\pi$-pulse pairs used for each momentum kick in the fast gates. The initial motional state is a thermal state with average occupation $\bar{n}=0.1$ for each mode. The red triangles in (a) represent a target gate fidelity to achieve a total simulation fidelity of 70\%, and in (b) represent the target time for a single UMQ gate to prevent significant trap heating effects during a simulation, given a heating rate of 10~s$^{-1}$. The purple triangles in (b) are an estimate of the gate time for a M\o lmer-S\o rensen gate, assuming a two-qubit gate time of 50~\si{\micro\second}.}\label{fig:1}
\end{figure}

The results of this optimisation are shown in Fig.~(\ref{fig:1}). An estimate of total fidelity for the UMQ gate on the trap is obtained by multiplying the fidelities of all the constituent fast gates. This should give a lower bound on the true UMQ fidelity, as this method assumes gate errors are perfectly correlated. We also calculated the threshold UMQ gate error beyond which a simulation fidelity of 70\% cannot be achieved (red triangles). This simulation corresponds to the Hubbard model used as an example in Lamata et al\cite{Lamata2013}, with lattice size corresponding to $L/2$ where $L$ is the ion number. We see that the achieved fidelities decay with the size of the trap, which is expected because larger traps have a more complex motional phase space, and because more two-qubit gates are needed to construct a UMQ gate in a larger trap. However, we do see very high fidelity overall, even for larger traps. Assuming perfect laser control and no trap heating or dephasing, the $n=50$ fast gate, corresponding to a minimum laser repetition rate of around 10~GHz, achieves a UMQ gate fidelity of 99.998\% in a 40-ion trap. This would correspond to fidelity of around 95\%, without including error correction, for 10 Trotter steps of a 20-site fermionic lattice simulation, requiring 2680 UMQ gates. UMQ gates with $n=50$ or more are capable, as shown in Fig.~(\ref{fig:1}a), of exceeding the error threshold required for a 20-site simulation without error correction with 70\% fidelity.

Current repetition rates for applicable pulsed laser systems are up to 300~MHz, with small modifications required to increase repetition rates to approximately 1~GHz \cite{Huss16pc,Johnson2015}. This is still well below the 10~GHz required for $n=50$ fast gates, but work is in progress on a pulsed laser system for $^{40}$Ca$^+$ control with a 5 GHz repetition rate \cite{pCommDaniel}. 

In Fig.~(\ref{fig:1}b), we analyze the dependence on time of our fast gate scheme. Again focusing on the $n=50$ case, the time required for a 40-ion UMQ gate is 17.9 \si{\micro\second}, comparable with the trap period of 10 \si{\micro\second} for this trap. Furthermore, the trap period does not place any fundamental limit on the speed of fast gates as it does with MS gates. 
In general, an MS gate for N ions will take a time $\sqrt{N}t_1$, $t_1$ being the time it takes for 1 ion. This can be seen from the expression of the gate's phase in terms of experimental parameters, $\theta \propto (\frac{\Omega \eta}{\delta})^2$. Here $\Omega$ is the Rabi frequency, which does not depend on the number of ions and is limited by the laser power, $\eta$ is the Lamb-Dicke parameter, which for $N$ ions is given by $\eta_1/\sqrt{N}$, where $\eta_1$ is the single-ion Lamb-Dicke parameter, when the COM vibrational mode is considered. The detuning $\delta$ sets the gate time to $t=1/\delta$. To compensate for the $1/N$ decay rate of $\theta$ with the number of ions, the detuning has to be decreased accordingly, $\delta \rightarrow \delta/{\sqrt{N}}$, which increases the time of the MS gate as $t \rightarrow \sqrt{N}t$. A two-qubit MS gate is typically around 50 \si{\micro\second} long~\cite{Benhelm2008}, thus a 40-ion gate takes an order of magnitude more time than our UMQ gate implementation with $n=50$, as shown in the figure. Higher available repetition rates would lead to even more significant gains, as seen in the $n=100$ case. The lowest heating rates in current traps are on the order of 10 s$^{-1}$, which limits the total simulation time to around 100 ms. This places a corresponding maximal threshold on gate time, shown in Fig. (\ref{fig:1}b). It can be seen that fast gates are below this threshold, and therefore that they would allow for a simulation outperforming classical computers within the trap coherence time.

A large scale quantum simulation with MS gates sits on the verge of current motional coherence times, and should be possible with coherence times expected in the near future. However, from the time scaling of MS (scaling as $\sqrt{N}$) and UMQ (linear in $N$) gates, it can be seen that UMQ gates would perform faster for a number of ions on the order of 1000 or below. This is far beyond the expected capacity of linear traps, and for this number of ions more creative ion arrays will be required, most likely composed of linear traps where the UMQ gates prove to be the most practical.

\begin{figure}[h]
\begin{center}

\includegraphics[width=0.5\textwidth]{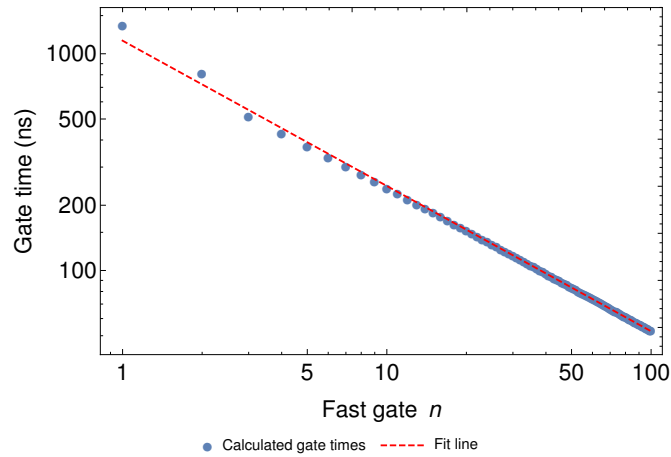}\
\caption{Gate times for FRAG fast gates on a two-ion trap with varying $n$. The points represent the numerical solutions for each $n$, and the dashed line is a least-squares fit to the points with the equation $T_G=1.149\ \si{\micro\second}\times n^{-0.6704}$, where $T_G$ is gate time.}\label{fig:moment}
\end{center}
\end{figure}

Finally, to achieve a better understanding of the interplay between laser repetition rate, momentum (i.e. fast gate $n$) and gate time, we numerically optimised fast gates using the FRAG scheme with $n=1$ to $n=100$ for our two-ion trap. The resulting gate times are shown in Figure (\ref{fig:moment}).

A simple least-squares fit to this data reveals that gate time scales with $n$ (and thus momentum) approximately as $n^{-2/3}$. This agrees very well with theoretical results for the GZC pulse scheme \cite{Garcia-Ripoll2003} and as our UMQ gates are constructed using fast gates, we expect the UMQ gate time to follow the same scaling.

To understand the relationship between laser repetition rate $f_r$ and gate time, we can use the approximation $n=T_Gf_r$, which works very well for fast gates in this regime \cite{Bentley2015}. We can combine this with the $n^{-2/3}$ scaling of gate time with the number of pulse pairs to arrive at
\begin{equation}
T_G\propto f_r^{-2/5}.
\end{equation}
The time taken for the entire simulation should scale according to this power law as achievable laser repetition rates increase.

\subsection*{Imperfection analysis}

There is a clear trend towards higher fidelity and lower gate time with higher momentum in the results shown in Fig.~(\ref{fig:1}), which means the development of faster pulsed lasers would lead to an even more favourable comparison with the MS gate. We have shown that under ideal conditions, a very high-fidelity UMQ gate can be performed using fast gates. However, to consider the experimental feasibility of performing simulations of interacting spins and fermions using fast gates, we analyse the effects of potential sources of error on the gates. In particular, errors in the energy of the resonant laser pulses (i.e. $\int_0^\tau\Omega(t)\neq\pi$, where $\Omega(t)$ is the Rabi frequency and $\tau$ is the pulse length) have been shown to affect gate fidelity significantly - it is expected that pulse energy fluctuations will be the most significant source of error for a fast gate \cite{Bentley2013}. We also consider dynamical effects of decoherence during the gate evolution.

To explore the effect of an error in $\pi$ pulse area on a gate applied to an adjacent pair of ions in a large trap, we employed a method relying on the assumption that the motional mode dynamics are separable during the gate, a valid assumption for small $\pi$ pulse errors in harmonic trapping potentials. This enables computationally feasible simulation of the gate evolution, allowing simple calculation of a representative-state fidelity for various levels of $\pi$ pulse error \cite{privcom}. Instead of using a low-temperature thermal state, requiring an inefficient simulation of the density matrix, the initial motional state is chosen as the first excited number state for each motional mode, allowing simulations with state vectors. The representative-state fidelity is
\begin{equation}
F=\left|\bra{\psi_0}U_{re}^\dagger U_{id}\ket{\psi_0}\right|^2,
\end{equation}
where $\ket{\psi_0}$ is the state $\frac{1}{\sqrt{2}}(\ket{00}+\ket{01}+\ket{10}+\ket{11})$, chosen to give an even superposition of the two-qubit computational basis states.

\begin{figure}[h]
\includegraphics[width=1\textwidth]{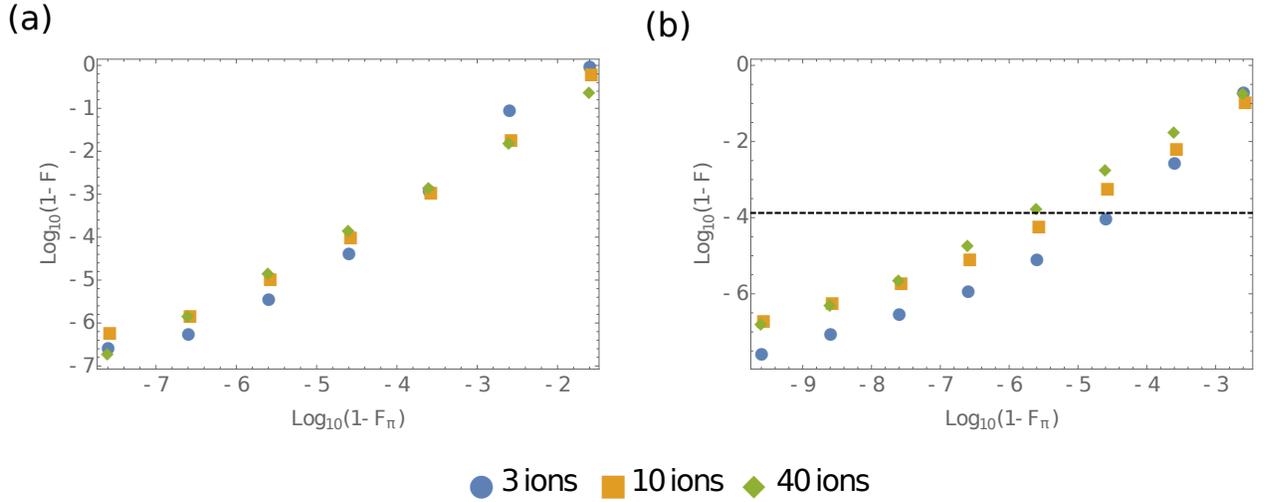}
\caption{A plot of infidelity for (a) a two-qubit $n=50$ fast gate and (b) a full UMQ gate against the rotational infidelity in each $\pi$ pulse, $F_\pi$. The error in (b) is estimated assuming the errors in successive fast gates are uncorrelated, and the dashed line represents a threshold per-gate error to achieve 70\% fidelity for the full simulation.}\label{fig:2}
\end{figure}

We simulated two-qubit fast gates on the outermost adjacent pair of ions (ions 1 and 2) in traps of varying size. The results of these simulations are shown in Fig.~(\ref{fig:2}a). Simulations used square pulses with a Rabi frequency given by $\Omega=\frac{\xi\pi}{2\tau}$, where $\tau$ is the pulse time such that $\xi=1$ gives a perfect $\pi$ pulse. Results are plotted against rotational infidelity $1-F_\pi$, where $F_\pi=\left|\bra{\psi_0} U_{\pi re}^\dag U_{\pi id}\ket{\psi_0}\right|^2$, and $U_{\pi re,id}$ are the real and ideal unitaries for the $\pi$ pulse. For this square pulse this rotational infidelity is approximately $\frac{\pi^2}{4}(1-\xi)^2$ \cite{privcom,Torosov2011,Ivanov2011}. Pulses within each pulse pair are assumed to have a relative phase of $\pi$ - this means there is no residual population transfer between the $\ket{0}$ and $\ket{1}$ states, even with imperfect pulse area, helping to reduce gate error from pulse imperfections \cite{Huss16pc, privcom}. Such a phase shift would be easily possible in an experiment using simple optics.

We also simulated a full UMQ gate on a three-ion trap with low-momentum ($n=2$) fast gates and imperfect $\pi$ pulses. Comparing the infidelity of the entire UMQ operation to that of a single fast gate with the same $\pi$ pulse imperfection, it was determined that errors due to $\pi$ pulse imperfections in successive fast gates are uncorrelated, and thus this type of error scales with the number of gates applied as $\sqrt{N}$. This allowed estimation of the error for a complete UMQ gate with imperfect $\pi$ pulses, shown in Fig.~(\ref{fig:2}b). Given this estimation, for rotational infidelity over $10^{-6}$, the gate infidelity is too high to implement a 40-ion simulation with an estimated fidelity above 70\%. Thus, we can place this limit on $\pi$ pulse errors for fast gates to be useful for a quantum simulation outperforming current classical computers without error correction. Proposals exist for composite pulse schemes which can produce extremely high fidelities, robust to the type of intensity instability or miscalibration that would damage a square pulse \cite{Torosov2011,Ivanov2011}. The error could also be reduced with the use of chirped laser pulses to perform robust, adiabatic population transfers \cite{Malinovsky2001}. The current state of the art for single-qubit operation error is approximately 10$^{-6}$ with microwave pulses \cite{Harty2014,Brown11PRA}, which compares favourably with the limit for pulse errors in this work, but it should be noted that using microwave pulses results in far slower gate times ($\sim$\si{\micro\second}) than ultrafast laser pulses. Single-qubit rotations have been demonstrated with ultrafast laser pulses \cite{Camp10PRL}, with error of around 0.01 and gate times of tens of \si{\pico\second}, meaning that the required single-qubit gate errors are yet to be achieved experimentally in the relevant regime. We expect single-qubit operation error to be similar to the achievable pulse errors, as single-qubit operations are usually performed with single or small numbers of pulses.

We also considered the effect of decoherence during fast gate evolution, in order to place a limit on the rate of decoherence processes to maintain the high fidelity of fast gates. In this case, a full simulation is not feasible for a 40-ion trap, so our results are inferred from simulations of decoherence effects on a low-momentum ($n=2$) fast gate on a two-ion trap. The full evolution of this gate was simulated, assuming the momentum kicks to be approximately instantaneous and thus unitary, and solving a master equation using a Monte Carlo trajectory approach during the free evolution periods of the gate.

We simulated the effects of trap heating and dephasing separately. The state of the trap evolves in the absence of laser light according to the master equations
\begin{equation}
\dot{\rho}=-\frac{i}{\hbar}[H,\rho]+ \Gamma_{\rm h} \sum_p (\mathcal{D}[a_p]+\mathcal{D}[a^\dagger_p])\rho ,
\end{equation}
\begin{equation}
\dot{\rho}=-\frac{i}{\hbar}[H,\rho]+ \Gamma_{\rm d} \sum_k \mathcal{D}[\sigma_z^k]\rho .
\end{equation}
Here, $\Gamma_{\rm h}$ and $\Gamma_{\rm d}$ are trap heating and dephasing rates, respectively, while $\sum_p$ represents a sum over motional modes and $\sum_k$ a sum over ions. Here, $\mathcal{D}[O]\rho\equiv O\rho O^\dag-\frac{1}{2}(O^\dag O\rho+\rho O^\dag O)$, and the Hamiltonian is given by $H=\sum_p \hbar\nu_pa^\dagger_pa_p$, where the vacuum energy has been set to 0 with a gauge transformation. We have assumed an infinite temperature reservoir for trap heating, a valid model for a randomly fluctuating electromagnetic field~\cite{Haffner2008}. Fidelity is again calculated using a representative state starting in the motional ground state. We trace out the motional state to calculate the computational fidelity after decoherence has been included in the simulations.

\begin{figure}
\includegraphics[width=1\textwidth]{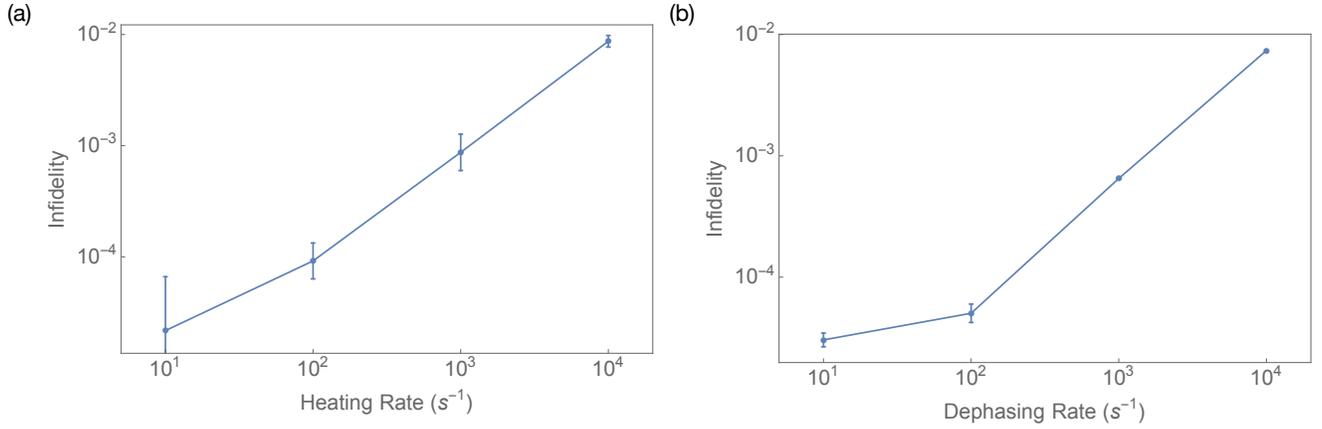}
\caption{Infidelity of an $n=2$ two-qubit fast gate in a two-ion trap under (a) trap heating and (b) dephasing.}
\label{fig:3}
\end{figure}

The results of these simulations are shown in Fig.~\ref{fig:3}. The infidelity calculated under trap heating shows that heating during gate evolution has a dynamical effect on the gate, damaging the phase accumulation required to implement the ideal unitary. The fidelities shown here are lower than that of a gate acting on a hotter initial state, meaning that the decoherence must be affecting the evolution of the gate. This can be understood physically as the heating changing the motional phase space trajectory of the gate. The effect of dephasing is similar to that of trap heating. The physical reasoning behind dephasing affecting the gate is more obvious, as dephasing directly introduces classical uncertainty in the phase of the ions' internal state, necessarily resulting in a loss of gate fidelity.

We can extend the conclusions drawn from these results to larger traps. If phonon absorption significantly damages the gate, we must work in a regime where a phonon absorption is quite unlikely during any operation. For a 50 ms simulation, this means that the heating rate is limited to approximately 10 s$^{-1}$. This would correspond to an absorption probability of 0.39 during the simulation, suggesting a simulation fidelity of 61\% in the absence of other errors. Current traps are typically capable of heating rates of this order of magnitude \cite{mckay14,Goodwin16,Bruzewicz2015}, with heating rates as low as 0.33 s$^{-1}$ demonstrated with cryogenic traps \cite{Niedermayr2014}.


\subsection*{Effect of higher-order terms in Coulomb interaction}
We also considered the effect of the anharmonicity of the Coulomb interaction on the fidelity of fast gates. Typically, the Coulomb interaction is truncated at second order when treating the motion of ions in a linear trap, in order to allow representation of the ions' motional states in terms of harmonic normal modes. This greatly simplifies calculations and was critical in enabling computationally efficient numerical searches for high-fidelity fast gates, but it is possible that this approximation neglects significant contributions from higher-order terms in the interaction. This is especially true because fast gates couple much more strongly to the motion than other gate techniques, enabling their higher speed.

To confirm the usefulness of gate schemes found using the harmonic approximation, we performed a full simulation of the dynamics of a two-ion trap during an $n=50$ fast gate using XMDS \cite{XMDS}, including the full Coulomb interaction. Trap specifications were the same as those used in the rest of this work. Fidelity computed from this simulation was identical to the fidelity of the same gate when truncating the interaction to second order, up to numerical precision of the simulation (approximately $10^{-6}$). This confirms that anharmonicity of the interaction has no appreciable effect on fast gates in the momentum ranges suggested in this work.

\section*{Discussion}

Fast gates have been proposed as a superior two-qubit gate for trapped ion quantum information processing, being much faster than existing gates and robust to many sources of error. In this work, we have analysed the potential for fast gates to implement useful digital quantum simulations, enough to outperform a classical computer. We have also studied the limits on experimental setups to successfully implement high-fidelity operations using fast gates. High-performing gates require rotational fidelities on the order of 10$^{-6}$ and decoherence rates below 10 s$^{-1}$. If these requirements are satisfied, it appears that fast gates can significantly outperform MS gates in terms of gate time, and can be implemented with very high fidelity if high-repetition-rate lasers are available~\cite{Huss16pc,Johnson2015}. Therefore, fast gates hold significant promise for the future of trapped ion quantum simulation and quantum information processing.

\section*{Acknowledgements}

We acknowledge support from Spanish MINECO/FEDER FIS2012-36673-C03-02 and FIS2015-69983-P; Ram\'on y Cajal Grant RYC-2012-11391; UPV/EHU Project EHUA14/04, UPV/EHU UFI 11/55 and a UPV/EHU PhD grant, Basque Government IT472-10 and IT986-16; PROMISCE, and SCALEQIT EU projrects. We also acknowledge support from the Australian Research Council Centre of Excellence for Quantum Computation and Communication Technology (CE110001027) and the Australian Research Council Future Fellowship (FT120100291) as well as DP130101613.

\section*{Author Contribution Statement}
R.L.T. performed analytical and numerical analysis of fast gates and wrote the bulk of the manuscript. C.D.B.B. contributed to the original conception and writing of the work, as well as the analytical and numerical frameworks and discussion of results. J.S.P., L.L. and E.S. contributed the theory of digital quantum simulations with trapped ions. J.J.H. and A.R.R.C. assisted in translating the MS gate scheme into pairwise nearest-neighbour fast phase gates, SWAP gates and single-qubit operations and determining the scaling of that decomposition with ion number, as well as co-writing parts of the manuscript and editing. All authors contributed to the editing process.

\section*{Additional Information}
The authors declare no competing financial interests.

\end{document}